# Inverse square Lévy walk emerging universally in goal-oriented tasks

Shuji Shinohara[a,*], Daiki Morita[a], Nobuhito Manome[b,c], Ryota Hayashi[a], Toru Moriyama[d], Hiroshi Okamoto[b], Pegio-Yukio Gunji[e], and Ung-il Chung[b]

[a] School of Science and Engineering, Tokyo Denki University, Saitama, Japan

[b] Department of Bioengineering, Graduate School of Engineering, The University of Tokyo, Tokyo, Japan

[c] Department of Research and Development, SoftBank Robotics Group Corp., Tokyo, Japan

[d] Faculty of Textile Science, Shinshu University, Ueda, Japan

[e] Department of Intermedia Art and Science, School of Fundamental Science and Technology, Waseda University, Tokyo, Japan

**\* Corresponding author**

Email: s.shinohara@mail.dendai.ac.jp

Postal address: School of Science and Engineering, Tokyo Denki University, Ishizaka, Hatoyama-machi, Hiki-gun, Saitama 350-0394, Japan

# Keywords

Lévy walk, Cauchy walk, Brownian walk, Bayesian inference, power-law distribution, goal-oriented

# Author contributions

**Shuji Shinohara:** Conceptualization, formal analysis, methodology, software, writing, original draft preparation, and funding acquisition. **Daiki Morita:** Software, reviewing, and editing. **Nobuhito Manome:** Writing, reviewing, and editing. **Ryota Hayashi:** Software, review, and editing. **Toru Moriyama:** Writing, reviewing, and editing. **Hiroshi Okamoto:** Writing, reviewing, and editing. **Pegio-Yukio Gunji:** Writing, reviewing, editing, and supervision. **Ung-il Chung:** Writing, review, editing, supervision, and project administration.

*Abstract—*

The Lévy walk, a type of random walk in which the frequency of linear step lengths follows a power-law distribution, can be observed in the migratory behavior of organisms at various levels, from bacteria and T cells to humans. Generally, Lévy walks with power exponents close to two are observed, and the reasons for such behavior are unclear. This study aims to propose a simple model that universally generates inverse square Lévy walks (also called Cauchy walks) and to identify the conditions under which Cauchy walks appear. We demonstrate that Cauchy walks emerge universally in goal-oriented tasks. We use the term "goal-oriented" when the goal is clear. However, this can be achieved in different ways, which cannot be uniquely determined. We performed an online simulation in which an agent observed the data generated from a certain probability distribution in a two-dimensional space and successively estimated the central coordinates of that probability distribution. This may be viewed as the task of a predator predicting where prey is most likely to be acquired based on prey-sighting memories. The agent has a model of probability distribution as a hypothesis for data-generating distribution and can modify the model such that each time a data point is observed, thereby increasing the estimated probability of occurrence of the observed data. To achieve this, the center coordinates of the model must be moved closer to those of the observed data. However, in the case of a two-dimensional space, arbitrariness arises in the direction of correction of the center coordinates; that is, this task is goal-oriented. We analyze the behavior of two cases: a strategy that allocates the amount of modification randomly in the *x*- and *y*-directions, and a strategy that determines allocation such that movement is minimized. The results reveal that when a random strategy is used, the frequency of the movement lengths shows a power-law distribution with exponent two; that is, the Cauchy walk appears. In contrast, the Brownian walk appears when the minimum strategy is used. Thus, the presence or absence of the constraint of minimizing the amount of movement may be a factor that causes the difference between Brownian and Lévy walks.

# I. INTRODUCTION

Lévy walks have been observed in the migratory behavior of organisms at various levels, from bacteria and T cells to humans [1][2][3][4][5]. Lévy walks are a type of random walk in which the frequency of a linear step length $l$ follows the power-law distribution $P(l) \sim l^{-\mu}, 1 < \mu \leq 3$. Compared to the Brownian walk (where the frequency of step length $l$ is characterized by an exponential distribution $P(l) \sim e^{-\lambda l}$), which is a type of random walk, the Lévy walk is characterized by the occasional appearance of very long linear moves. Generally, Lévy walks with exponents close to two have been observed in the migratory behavior of organisms, and attention has been paid to why such patterns occur [1][6][7][8][9][10][11]. Hereafter, the inverse square Lévy walk with an exponent of two is also referred to as a Cauchy walk. The Lévy flight foraging hypothesis (LFFH) [12][13] states that if food is sparse and randomly scattered and predators have no information (memory) about the food, Lévy walks, as a random search, will be the optimal foraging behavior and will be evolutionarily advantageous. Until now, it has been considered that search efficiency is maximized for inverse square Lévy walks with an exponent of two in the LFFH [14]. However, recent studies have indicated that a Cauchy walk does not necessarily have maximum search efficiency in spaces of two or more dimensions, as this is true only under special conditions [15]. In contrast, Guinard and Korman proved that an intermittent Cauchy walk is an optimal search strategy in finite two-dimensional domains when the goal is to rapidly find targets of arbitrary sizes [16]. Debates about natural conditions and search methods that make the Cauchy walk optimal are ongoing.

Although the observation of the Levy walk has been attributed to the execution an optimal search strategy for sparsely and randomly distributed resources, this interpretation has not always been generally accepted [17][18][19]. From a different perspective than LFFH, Abe hypothesized that the functional advantages of the Lévy walk arise from the critical phenomena of the system and demonstrated that Lévy walks appear near a critical point between stable synchronous and unstable asynchronous states [20]. The reason Lévy walks are observed in living organisms is that near the critical point, the range of inputs from which information can be discriminated is larger, giving the organism the flexibility to switch between searching for nearby and new distant locations depending on the input. In Abe's model, the Lévy walk appears near the critical point. However, it is not a Cauchy walk with an exponent of two. Sakiyama developed an algorithm that generates Cauchy walks as a result of one walker's decision-making process [10]. Huda et al. demonstrated that the trajectories of metastatic cells display the Lévy walks, while non-metastatic cancerous cells perform simple diffusive movements [11].

The main objective of this study was to propose a simple model that universally generates Cauchy walks and to identify the conditions under which Cauchy walks appear. Specifically, we show that the Cauchy walk emerges in goal-oriented tasks and analyze the reasons for its emergence. We use the term "goal-oriented" in the following sense: First, consider a function $z = f(x, y)$ consisting of two variables, $x$ and $y$. Additionally, we aim to shift $x$ and $y$ to change $z$ by a small amount $\Delta z$. The task here is to obtain $\Delta x$ and $\Delta y$ such that $\Delta z = f(x + \Delta x, y + \Delta y) - f(x, y)$. The amount of movement of $x$ required to realize the objective can be approximated as $\Delta x \approx \Delta z \frac{\partial x}{\partial z}$ using partial differentiation. Similarly, the amount of movement of $y$ can be approximated as $\Delta y \approx \Delta z \frac{\partial y}{\partial z}$. Thus, changing $z$ by $\Delta z$ can be realized by moving only one of $x$ or $y$, or by moving both $x$ and $y$ in any allocation. This is when arbitrariness arises. One strategy may be to equally allocate $\frac{\Delta z}{2}$ in both the $x$- and $y$-directions as $\Delta x \approx \frac{\Delta z}{2} \frac{\partial x}{\partial z}$ and $\Delta y \approx \frac{\Delta z}{2} \frac{\partial y}{\partial z}$. Another strategy may be to determine the allocation of $\Delta x$ and $\Delta y$ to minimize the amount of movement $l = \sqrt{\Delta x^2 + \Delta y^2}$, as in the steepest descent method. However, these strategies are not essential to achieving the purpose of changing $z$ to $z + \Delta z$. Thus, we use the term "goal-oriented" when the goal is clear. However, this can be achieved in multiple ways and cannot be uniquely determined.

In this study, we considered a two-dimensional normal distribution as a specific form of *z*. Furthermore, we considered an online estimation task in which data that were randomly generated based on a certain probability distribution were observed individually and the center (mean), one of the distribution parameters, was estimated. Such tasks are common in machine learning and include Bayesian inference and online clustering [21]. Additionally, it can be viewed as the task of a predator predicting where prey is most likely to be acquired based on memories of prey sightings at various locations.

In this task, the estimates of the center coordinates must be revised sequentially, based on the observed data. To address this, we use Bayesian inference with a discount rate (BID) [22][23][24]. In a BID, several probability distributions (called models) are first prepared as hypotheses for the probability distribution that generates the data. Each time data are observed, the degree of confidence in each hypothesis is updated by quantitatively evaluating the fit between each hypothetical model and the observed data. Finally, the best hypothesis is narrowed to a single one based on confidence levels. BID also modifies the hypothesis model by shifting the center (mean) to increase the estimated probability of the occurrence of the observed data. Thus, increasing the probability of data occurrence may be equivalent to predators predicting that prey will be more likely acquired at the same location. In this task, although the BID defines the degree to which the probability of occurrence of the observed data is increased, that is, $\Delta z$, it does not determine the direction of movement of the center coordinates, and the allocation of correction in the *x*- and *y*-directions is arbitrary. Thus, the BID is goal-oriented.

In this study, we analyzed the behavior of two cases: a strategy that allocates the amount of modification randomly in both the *x*- and *y*-directions as $\Delta x \approx \beta \Delta z \frac{\partial x}{\partial z}$ and $\Delta y \approx (1-\beta) \Delta z \frac{\partial y}{\partial z}$, respectively, where $0 \leq \beta \leq 1$ and $\beta$ are determined randomly, and a strategy that determines the allocation such that movement $l = \sqrt{\Delta x^2 + \Delta y^2}$ is minimized. The former can be considered a non-minimum modification strategy because it does not minimize the amount of movement. The amount of modification *l* in the non-minimum strategy is longer than that in the minimum strategy because it does not have the constraint of minimizing the movement. The results of the analysis revealed that when a non-minimum strategy is used, the frequency of movement *l* shows a power-law distribution with exponent two; that is, a Cauchy walk appears. By contrast, a Brownian walk appears when the minimum strategy is used.

Thus, the presence or absence of the constraint of minimizing movement in goal-oriented tasks may be one factor causing the difference between Brownian and Lévy walks. A lack of means or inability to detect the shortest path may have contributed to the emergence of Lévy walks for this task instead of their inherent advantages.

## II. METHODS

### 2.1 Online distribution estimation task

Consider an online estimation task in which an agent observes randomly generated data based on a certain probability distribution in a two-dimensional space, one at a time, and estimates the mean (center) of the distribution (Fig. 1).

In this study, we considered two types of data-generating distributions: normal and circular. The task of estimating the parameters of the generative distribution from the data is common in machine learning and may be considered the task of estimating the true values from measurements that contain errors. Additionally, this task can be viewed as predicting the most likely location for prey acquisition based on memories of previous prey sightings at various locations.

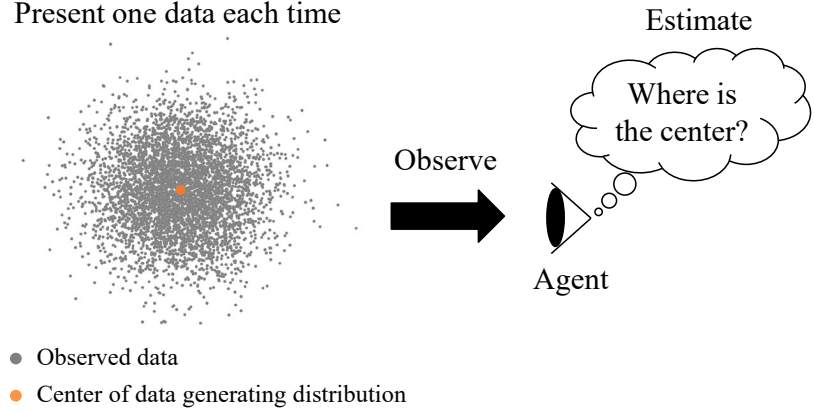

- Observed data
- Center of data generating distribution

**Figure 1. Online estimation task overview. The agent observes randomly generated data based on a certain probability distribution one by one each time and successively estimates the central coordinates of the distribution.**

## 2.2 Mean value estimation algorithms

### 2.2.1 Exponential moving average (EMA)

We describe one of the most basic methods of online estimation, the exponential moving average (EMA), which is generally used in online learning algorithms such as the expectation-maximization algorithm [25] and learning vector quantization [26].

If the mean estimate at time $t$ is $\mu^t$ and the observed data is $d^t$, the estimate at the next time $t+1$ is expressed as

$$\mu^{t+1} = (1-\alpha)\mu^t + \alpha d^t = \mu^t + \alpha(d^t - \mu^t) = \mu^t + \Delta\mu. \qquad (1)$$

Here, $0 \leq \alpha \leq 1$ is the discount rate and $\Delta\mu = \alpha(d^t - \mu^t)$. This formula implies that the estimate at the next time is expressed as a weighted average of the current estimate and observed data; that is, each time data are observed, the estimate is moved closer to the data by $\Delta\mu$. However, reducing the distance between them by $\Delta\mu$ does not necessarily imply moving the estimate by the same amount. As can be seen by drawing concentric circles as auxiliary lines, if the goal is to reduce the distance between the estimate and data by $\Delta\mu$, the estimate can be moved to anywhere on one of the inner concentric circles. Moreover, EMA is an algorithm that moves the estimate closer to the data at the shortest distance (Fig. 2).

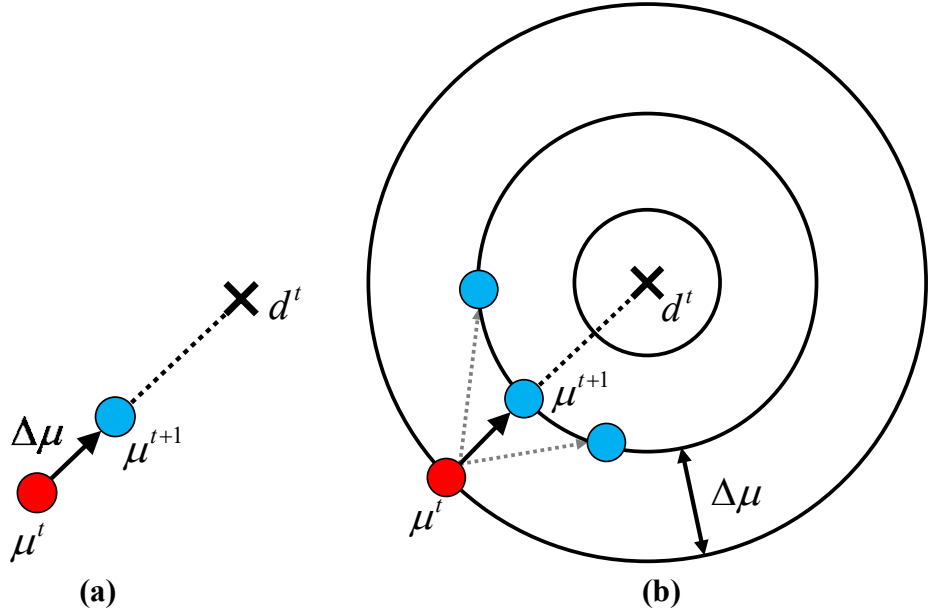

**Figure 2. Overview of EMA. (a) EMA is an algorithm that moves the estimate close to the observed data. (b) If the goal is simply to reduce the distance between the estimate and the data by $\Delta\mu$, the estimate can be moved anywhere on the one inner concentric circle. Conversely, EMA is an algorithm that moves the estimate close to the data at the shortest distance.**

If we focus on recursion, Eq. (1) can be transformed as

$$\mu^{t+1} = \alpha d^t + (1-\alpha)\alpha d^{t-1} + (1-\alpha)^2 \alpha d^{t-2} + \cdots + (1-\alpha)^n \alpha d^{t-n} + \cdots + (1-\alpha)^{t-1} \alpha d^1 + (1-\alpha)^t \mu^0$$
$$= \sum_{n=0}^{t}(1-\alpha)^n \alpha d^{t-n} + (1-\alpha)^{t+1}\mu^0. \tag{2}$$

When $\alpha = 0$, $\mu^{t+1} = \mu^0$ and the estimated values remain unchanged from the initial values. When $\alpha = 1$, $\mu^{t+1} = d^t$ and the observed data at that time become the estimated value next time. In the case of $0 < \alpha < 1$, the effect of the observed data in the distant past on the current estimate weakens exponentially.

Additionally, we consider a special case of EMA in which all observations are captured equally. The mean estimate at time $t+1$ is computed as

$$\mu^{t+1} = \frac{d^0 + d^2 + \cdots + d^t}{t+1}. \quad (3)$$

Similarly, the estimate of time $t$ can be calculated as

$$\mu^t = \frac{d^0 + d^2 + \cdots + d^{t-1}}{t}. \quad (4)$$

Focusing on recursion, the estimate for time $t+1$ can be written as

$$\mu^{t+1} = \frac{t}{t+1}\mu^t + \frac{1}{t+1}d^t = \left(1 - \frac{1}{t+1}\right)\mu^t + \frac{1}{t+1}d^t. \quad (5)$$

This corresponds to the case of the EMA, where the discount rate $\alpha = \frac{1}{t+1}$ decays with time. The more data accumulated, the smaller the discount rate, and the less weight is given to newly captured observation data.

Making the estimates closer to the observed data using Eq. (1) implies that information from the current observation is reflected in the estimate. In contrast, as shown in Eq. (2), the current estimates incorporate information from historical observations. Moreover, the strategy of minimizing the amount of movement used in the EMA involves incorporating information from new observations into the estimates, while minimizing the loss of information from past data as much as possible.

## 2.3 Bayesian inference with discount rate (BID)

### 2.3.1 Bayesian inference

In Bayesian inference, several hypotheses are formulated for the estimation target. Furthermore, evidence (data generated from the estimation target) is observed, and the degree of confidence in each hypothesis is updated by quantitatively evaluating the fit between each hypothesis and the observed data. Finally, the best hypothesis is narrowed to one based on the confidence level.

This study deals with discrete Bayesian inference using a finite number of hypotheses. First, several hypotheses $h_k$ are defined, and the model for each hypothesis (the generated distribution of data $d$) is prepared in the form of a conditional probability $P(d|h_k)$. This conditional probability is called the likelihood when the data are fixed and considered a function of the hypotheses $h_k$. The initial value $P^0(h_k)$ of the confidence level for each hypothesis is prepared as a prior probability.

If the confidence level for the hypothesis $h_k$ at time $t$ is $P^t(h_k)$ and data $d^t$ is observed, the posterior probability $P^t(h_k|d^t)$ is computed using Bayes' theorem as

$$P^t(h_k|d^t) = \frac{P^t(h_k)P(d^t|h_k)}{P^t(d^t)}, \quad (6)$$

where $P^t(d^t)$ is the marginal probability of the data at time $t$ and is defined as

$$P^t(d^t) = \sum_k P^t(h_k)P(d^t|h_k). \quad (7)$$

Additionally, the posterior probability is read into the prior probability (confidence level) next time by the following Bayesian update:

$$P^{t+1}(h_k) \leftarrow P^t(h_k|d^t). \quad (8)$$

Summing up Eqs. (6) and (8), we obtain

$$P^{t+1}(h_k) \leftarrow \frac{P^t(h_k)P(d^t|h_k)}{P^t(d^t)}. \quad (9)$$

The estimation proceeds by updating the confidence level for each hypothesis by Eq. (9) each time the data is observed. The sum of the confidence levels for each hypothesis satisfies $\sum_k P^t(h_k) = 1$. Although the confidence level changes, the model $P(d|h_k)$ for each hypothesis is invariant throughout time evolution.

## 2.4 Introduction of a discount rate

In general, for an accurate estimation of a target, large amounts of data (information) should be obtained. However, this only occurs in stationary environments. Observation data from the distant past should be discarded when the target of estimation changes dynamically for precise estimation. Additionally, a few situations can arise in which large amounts of data cannot be acquired simultaneously. To address such situations, this study defines BID as [22][23][24].

$$P^{t+1}(h_k) \leftarrow \left[(1-\alpha)P^t(h_k|d^t)^m + \alpha P^t(d^t|h_k)^m\right]^{\frac{1}{m}} \quad (10)$$

$$P^{t+1}(d^t|h_k) \leftarrow \left[(1-\alpha)P^t(d^t|h_k)^m + \alpha P^t(h_k|d^t)^m\right]^{\frac{1}{m}} \quad (11)$$

Here, $\left[(1-\alpha)x^m + \alpha y^m\right]^{\frac{1}{m}}$ represents the generalized weighted average of $x$ and $y$; $0 \leq \alpha \leq 1$ denotes the weighting of $x$ and $y$; and $-\infty \leq m \leq \infty$ denotes the average.

This expression represents the arithmetic mean when $m = 1$ and the harmonic mean when $m = -1$. The case of $m = 0$ cannot be defined because it involves division by zero. However, the limit $m \to 0$ represents the geometric mean. In this study, we considered the case where $m = 1$. If $\alpha = 0$, Eq. (10) agrees with Eq. (8). Similarly, if $\alpha = 0$, Eq. (11) becomes $P^{t+1}(d^t|h_k) \leftarrow P^t(d^t|h_k)$, and the model remains unchanged. Thus, when $\alpha = 0$ is used, the BID is consistent with Bayesian inference.

This method is an inverse Bayesian inference algorithm [27][28][29] proposed by Gunji et al. in which a symmetry bias is incorporated into Bayesian inference to reproduce human causal induction [22]. Equation (11) shows that for $\alpha > 0$, the hypothetical model is modified such that the symmetry $P^t(d^t|h_k) = P^t(h_k|d^t)$ is satisfied. If we focus on recursion, as in the case of the EMA, Eq. (11) can be transformed as

$$P^{t+1}(d^t|h_k) = \alpha P^t(h_k|d^t)^t + (1-\alpha)\alpha P^{t-1}(h_k|d^{t-1}) + (1-\alpha)^2 \alpha P^{t-2}(h_k|d^{t-2}) + \cdots + (1-\alpha)^n \alpha P^{t-n}(h_k|d^{t-n}) +$$

$$\cdots + (1-\alpha)^{t-1} \alpha P^1(h_k|d^1) + (1-\alpha)^t P^0(d^0|h_k)$$

$$= \sum_{n=0}^{t}(1-\alpha)^n \alpha P^{t-n}(h_k|d^{t-n}) + (1-\alpha)^{t+1} P^0(d^0|h_k)$$

. (12)

In the case of $0 < \alpha < 1$, more weight is assigned to the more recent posterior probabilities, which is reflected in the model. Thus, the model changes based on the observed data. Therefore, in the BID, a subscript "$t$" is added such as $P^t(d^t|h_k)$ thereafter.

In this study, the observed data $d$ are considered to be two-dimensional vectors, and the model $P^t(d|h_k)$ for each hypothesis is a two-dimensional normal distribution as

$$P^t(d|h_k) = P^t(d|\mu_k^t, \Sigma_k^t) = \frac{1}{\sqrt{(2\pi)^2|\Sigma_k^t|}} \exp\left(-\frac{1}{2}(d-\mu_k^t)^T(\Sigma_k^t)^{-1}(d-\mu_k^t)\right), \quad (13)$$

where $\Sigma_k$ is the covariance matrix and $\mu_k$ is the center (mean) of the model for hypothesis $k$. When the number of hypotheses is set to a finite number and a normal distribution is introduced as the hypothetical model, $P(h_k)$ and $P(h_k|d)$ represent probabilities. Additionally, $P(d|h_k)$ is a probability density function. Therefore, Eqs. (10) and (11) are corrected as follows based on studies [23]:

$$P^{t+1}(h_k) \leftarrow (1-\alpha)P^t(h_k|d^t) + \alpha \Delta dP^t(d^t|h_k), \quad (14)$$

$$\Delta dP^{t+1}(d^t|h_k) \leftarrow \alpha P^t(h_k|d^t) + (1-\alpha)\Delta dP^t(d^t|h_k), \quad (15)$$

where $\Delta d = \sqrt{(2\pi)^2|\Sigma_k^t|}$. By making this correction, $\Delta dP^t(d^t|h_k)$ takes a value in the range of $0 \leq \Delta dP^t(d^t|h_k) = \exp\left(-\frac{1}{2}(d-\mu_k^t)^T(\Sigma_k^t)^{-1}(d-\mu_k^t)\right) \leq 1$. If both sides of Eq. (15) are divided by $\Delta d$, it can be transformed as follows:

$$P^{t+1}\left(d^{\,t}\mid \mu_k^{t+1},\Sigma_k^{t+1}\right) \leftarrow \alpha \frac{P^t\left(\mu_k^t,\Sigma_k^t\mid d^{\,t}\right)}{\Delta d}+(1-\alpha)P^t\left(d^{\,t}\mid \mu_k^t,\Sigma_k^t\right). \quad (16)$$

In Eq. (16), if the observed data $d^t$ is fixed,
$P^t\left(d^{\,t}\mid \mu_k^t,\Sigma_k^t\right)=\frac{1}{\sqrt{(2\pi)^2\left|\Sigma_k^t\right|}}\exp\left(-\frac{1}{2}\left(d^{\,t}-\mu_k^t\right)^{\mathrm{T}}\left(\Sigma_k^t\right)^{-1}\left(d^{\,t}-\mu_k^t\right)\right)$ is considered a function of the covariance matrix $\Sigma_k^t$ and the mean $\mu_k^t$, that is, the likelihood. Hereafter, as we stated that a likelihood is present, $P^t\left(d^{\,t}\mid \mu_k^t,\Sigma_k^t\right)$ is denoted as $P^t\left(\mu_k^t,\Sigma_k^t\right)$ or $Z_k^t$.

## 2.5 Modification of the model

For $0<\alpha<1$, the model is modified based on Eq. (16). The model modification for hypothesis $k$ when the data $d^t$ is obtained is as

$$\begin{aligned}
\Delta Z_k^t &= Z_k^{t+1}-Z_k^t \\
&= \left[\alpha\frac{P^t\left(\mu_k^t,\Sigma_k^t\mid d^{\,t}\right)}{\Delta d}+(1-\alpha)P^t\left(\mu_k^t,\Sigma_k^t\right)\right]-P^t\left(\mu_k^t,\Sigma_k^t\right). \quad (17)\\
&= \alpha\left[\frac{P^t\left(\mu_k^t,\Sigma_k^t\mid d^{\,t}\right)}{\Delta d}-P^t\left(\mu_k^t,\Sigma_k^t\right)\right]
\end{aligned}$$

Furthermore, from Eq. (13):

$$\frac{\partial P}{\partial \mu}=\frac{\partial Z}{\partial \mu}=Z\cdot \Sigma^{-1}(d-\mu), \quad (18)$$

$$\frac{\partial P}{\partial \Sigma}=\frac{\partial Z}{\partial \Sigma}=Z\cdot\left[-\frac{1}{2}\Sigma^{-1}+\frac{1}{2}\left(\Sigma^{-1}(d-\mu)(d-\mu)^T \Sigma^{-1}\right)\right]. \quad (19)$$

Therefore, when the mean and variance modifications are sufficiently small, they can be approximated as

$$\Delta \mu_k^t \approx \Delta Z_k^t\frac{\partial \mu}{\partial Z_k^t}=\frac{\Delta Z_k^t}{Z_k^t\cdot\left(\Sigma_k^t\right)^{-1}\left(d^t-\mu_k^t\right)}, \quad (20)$$

$$\Delta\Sigma_k^t \approx \Delta Z_k^t \frac{\partial \Sigma}{\partial Z_k^t} = \frac{\Delta Z_k^t}{Z_k^t \cdot \left[ -\frac{1}{2}\left(\Sigma_k^t\right)^{-1} + \frac{1}{2}\left(\left(\Sigma_k^t\right)^{-1}\left(d^t - \mu_k^t\right)\left(d^t - \mu_k^t\right)^T \left(\Sigma_k^t\right)^{-1}\right)\right]}. \quad (21)$$

Because this study deals with the task of estimating the mean, the covariance matrix is fixed, and only the correction of the mean is dealt with thereafter.

Because Eq. (20) uses partial differentiation, the amount of correction must be small for the approximate equation to be valid. Therefore, an upper limit is set for $\Delta\mu_k^t$ such that it does not exceed the threshold value $\Delta_{max}$.

For convenience, we consider a case in which only one hypothesis exists. Even if multiple hypotheses are present, if the confidence level of one hypothesis becomes one and the hypothesis is narrowed down to one, the situation is the same as that in the case of only one hypothesis. In this case, the confidence level expressed in Eq. (14) is not updated, and only the model is modified based on Eq. (15) or (16). If we eliminate the subscript $k$, which identifies the hypothesis, and denote only the hypothesis as $h$, then $P(h) = P(h|d) = 1$ holds. Therefore, Eq. (15) can be rewritten as

$$\Delta dP^{t+1}\left(d^t \mid h\right) \leftarrow \alpha \cdot 1 + (1-\alpha)\Delta dP^t\left(d^t \mid h\right). \quad (22)$$

That is, $\Delta dP^{t+1}\left(d^t \mid h\right)$ is represented by a weighted average of one and $\Delta dP^t\left(d^t \mid h\right)$. As the hypothetical model is considered to have a normal distribution,

$$0 \leq \Delta dP^t\left(d^t \mid h\right) = \exp\left(-\frac{1}{2}\left(d^t - \mu^t\right)^T \left(\Sigma^t\right)^{-1}\left(d^t - \mu^t\right)\right) \leq 1 = \exp\left(-\frac{1}{2}\left(\mu^t - \mu^t\right)^T \left(\Sigma^t\right)^{-1}\left(\mu^t - \mu^t\right)\right).$$

Thus, $\Delta dP^{t+1}\left(d^t \mid h\right) > \Delta dP^t\left(d^t \mid h\right)$ i.e., $P^{t+1}\left(d^t \mid h\right) > P^t\left(d^t \mid h\right)$ holds for $d^t \neq \mu^t$, resulting in $P^{t+1}\left(d^t \mid h\right) - P^t\left(d^t \mid h\right) > 0$. This implies that when data $d^t$ is observed, the model should be modified to increase the probability of $d^t$. To increase the probability of occurrence, the mean estimate of $\mu^t$ must be made closer to the observed data $d^t$, as in the EMA. The specific modification amount $\Delta \mu^t$ can be calculated using Eqs. (17) and (20) as

$$\Delta \mu^t = \begin{pmatrix} \Delta \mu_x^t \\ \Delta \mu_y^t \end{pmatrix} \approx \begin{pmatrix} \Delta Z^t \frac{\partial \mu_x}{\partial Z} \\ \Delta Z^t \frac{\partial \mu_y}{\partial Z} \end{pmatrix}$$

$$= \frac{\Delta Z^t}{Z^t \cdot (\Sigma)^{-1}(d^t - \mu^t)}$$

$$= \alpha \left[ \frac{P^t(\mu^t, \Sigma \mid d^t)}{\Delta d} - Z^t \right] \frac{1}{Z^t \cdot (\Sigma)^{-1}(d^t - \mu^t)}$$

$$= \alpha \left[ \frac{1}{\Delta d} - Z^t \right] \frac{1}{Z^t \cdot (\Sigma)^{-1}(d^t - \mu^t)}$$

$$= \frac{\alpha}{\Delta d} \left[ 1 - \exp\left(-\frac{1}{2}(d^t - \mu^t)^{\mathrm{T}}(\Sigma)^{-1}(d^t - \mu^t)\right) \right] \frac{1}{\frac{1}{\Delta d} \exp\left(-\frac{1}{2}(d^t - \mu^t)^{\mathrm{T}}(\Sigma)^{-1}(d^t - \mu^t)\right)(\Sigma)^{-1}(d^t - \mu^t)}$$

$$= \alpha \frac{\left[ 1 - \exp\left(-\frac{1}{2}(d^t - \mu^t)^{\mathrm{T}}(\Sigma)^{-1}(d^t - \mu^t)\right) \right]}{\exp\left(-\frac{1}{2}(d^t - \mu^t)^{\mathrm{T}}(\Sigma)^{-1}(d^t - \mu^t)\right)} \cdot \frac{1}{(\Sigma)^{-1}(d^t - \mu^t)} \tag{23}$$

Here, the covariance matrix is fixed at $\Sigma$. In this study, for simplicity, $\Sigma$ is assumed to be a diagonal matrix and $\Sigma = \begin{pmatrix} \sigma^2 & 0 \\ 0 & \sigma^2 \end{pmatrix}$. Thus, the inverse of $\Sigma$ is $\Sigma^{-1} = \frac{1}{\sigma^4}\begin{pmatrix} \sigma^2 & 0 \\ 0 & \sigma^2 \end{pmatrix}$. Furthermore, we let

$d^t - \mu^t = \begin{pmatrix} d_x^t - \mu_x^t \\ d_y^t - \mu_y^t \end{pmatrix} = \begin{pmatrix} r_x^t \\ r_y^t \end{pmatrix} = r^t$. Then, Eq. (23) can be rewritten as

$$\Delta \mu^t = \begin{pmatrix} \Delta \mu_x^t \\ \Delta \mu_y^t \end{pmatrix} = \begin{pmatrix} \Delta Z^t \frac{\partial \mu_x}{\partial Z} \\ \Delta Z^t \frac{\partial \mu_y}{\partial Z} \end{pmatrix}$$

$$\approx \alpha \frac{\left[1 - \exp\left(-\frac{\|r^t\|^2}{2\sigma^2}\right)\right]}{\exp\left(-\frac{\|r^t\|^2}{2\sigma^2}\right)} \cdot \frac{1}{\frac{1}{\sigma^4}\begin{pmatrix} \sigma^2 & 0 \\ 0 & \sigma^2 \end{pmatrix}\begin{pmatrix} r_x^t \\ r_y^t \end{pmatrix}}. \quad (24)$$

$$= \alpha \sigma^2 \left[\exp\left(\frac{\|r^t\|^2}{2\sigma^2}\right) - 1\right]\begin{pmatrix} \frac{1}{r_x^t} \\ \frac{1}{r_y^t} \end{pmatrix}$$

When $\frac{\|r^t\|^2}{2\sigma^2}$ is near zero, equation (24) can be transformed using Maclaurin expansion as

$$\Delta \mu^t \approx \alpha \sigma^2 \left[\exp\left(\frac{\|r^t\|^2}{2\sigma^2}\right) - 1\right]\begin{pmatrix} \frac{1}{r_x^t} \\ \frac{1}{r_y^t} \end{pmatrix}$$

$$\approx \alpha \sigma^2 \left(1 + \frac{\|r^t\|^2}{2\sigma^2} - 1\right)\begin{pmatrix} \frac{1}{r_x^t} \\ \frac{1}{r_y^t} \end{pmatrix} \quad (25)$$

$$= \frac{\alpha}{2}\|r^t\|^2 \begin{pmatrix} \frac{1}{r_x^t} \\ \frac{1}{r_y^t} \end{pmatrix}$$

Equation (17) requires that the probability of occurrence of the observed data $d^t$ is corrected by $\Delta Z^t$. $\Delta \mu_x^t = \Delta Z^t \frac{\partial \mu_x}{\partial Z}$ implies that if the mean is shifted by $\Delta \mu_x^t$ in the *x*-direction, the probability of occurrence is corrected by $\Delta Z^t$; the same is true in the *y*-direction. Thus, $\Delta Z^t$ can be modified by shifting the mean in either the *x*- or *y*-direction or by mixing the two in arbitrary proportions. Moreover, arbitrariness arises in the direction and amount of mean correction (Fig. 3).

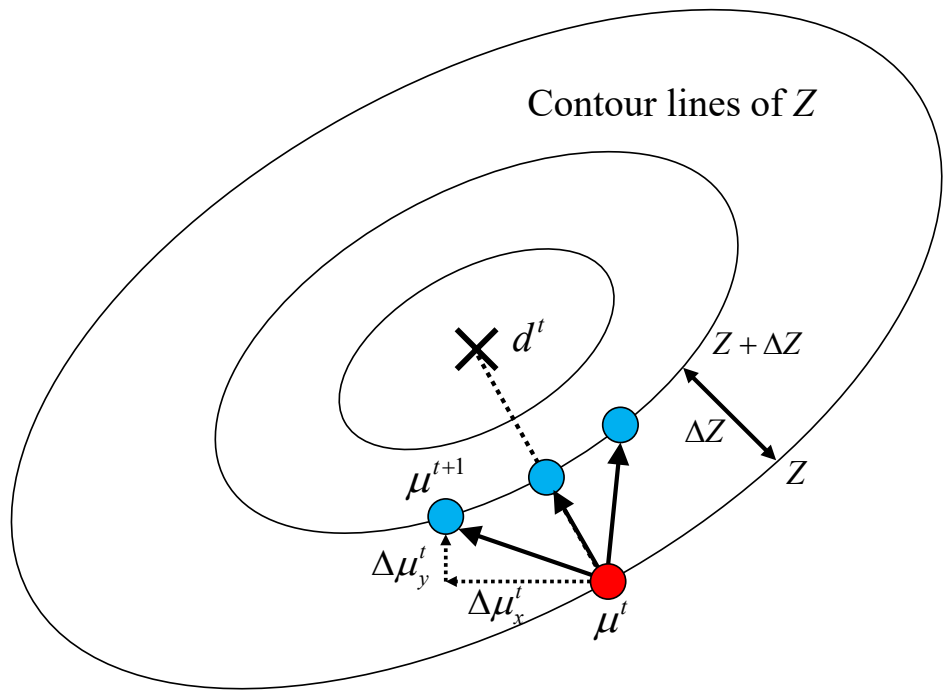

**Figure 3. Likelihood contours. If we want to increase the likelihood by $\Delta Z$, we can move the estimate $\mu$ to any position on the inner contour. Moreover, an infinite number of possible solutions are presented.**

Figure 3 shows a conceptual diagram of the likelihood $Z = P(\mu) = P(\mu \mid d^t, \Sigma) = \dfrac{1}{\sqrt{(2\pi)^2 |\Sigma|}} \exp\left(-\dfrac{1}{2}(d^t - \mu)^{\mathrm{T}} (\Sigma)^{-1} (d^t - \mu)\right)$ when the observed data $d^t$ and the covariance matrix $\Sigma$ are fixed in Eq. (13) and viewed as a function of the estimated value $\mu$. We proposed and compared two modification methods. The first is to make the modification amount $\Delta Z^t$ allocated randomly in both the *x*- and *y*-directions such as $\Delta \mu_x^t \approx \beta \Delta Z^t \dfrac{\partial \mu_x}{\partial Z^t}$ and $\Delta \mu_y^t \approx (1-\beta) \Delta Z^t \dfrac{\partial \mu_y}{\partial Z^t}$. Here $0 \leq \beta \leq 1$ and $\beta$ is determined randomly each time a modification is performed. The second is the same as the EMA to move the estimated value $\mu^t$ close to the observed data $d^t$ in the shortest distance, i.e., along the straight line connecting the current estimate $\mu^t$ and observed data $d^t$. Hereafter, the agents

using each strategy are referred to as non-Min and Min agents, respectively.

Here, with respect to partial differentiation, the agent's coordinate system (i.e., the directions of the x- and y-axes) can be arbitrary. Therefore, in this study, the directions of the *x*- and *y*-axes were defined randomly each time a modification was performed.

## 2.6 Parameter setting

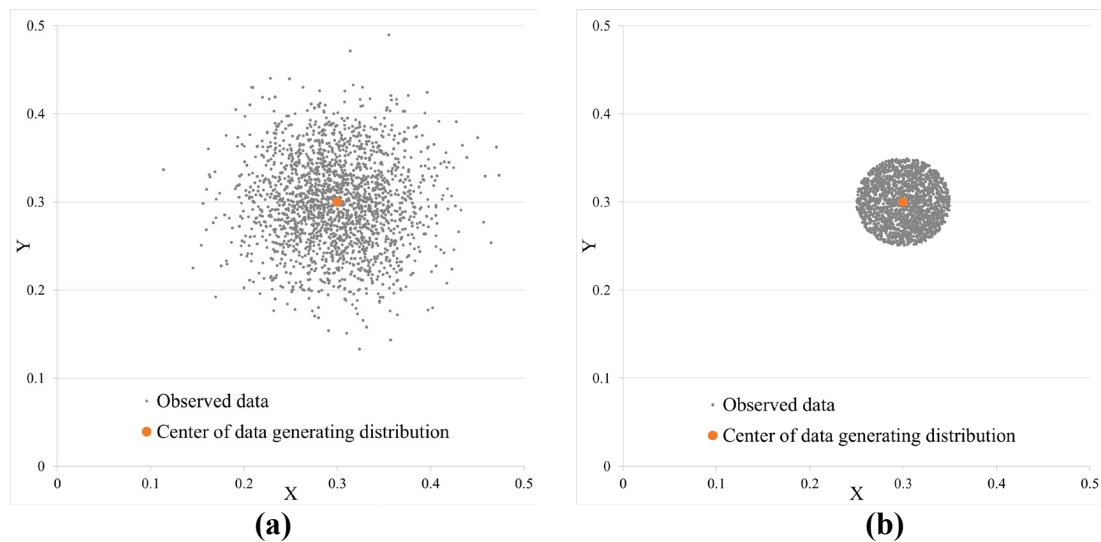

**Figure 4. Examples of observed data. (a) When the data-generating distribution is a normal distribution. (b) When the data-generating distribution is a circular uniform distribution.**

Two types of data-generating distributions were considered: normal and uniform circular distributions. The centers (means) of the normal and uniform distributions were set to $\begin{pmatrix} 0.3 \\ 0.3 \end{pmatrix}$. The covariance matrix of the normal distribution is set to $\begin{pmatrix} 0.0025 & 0 \\ 0 & 0.0025 \end{pmatrix}$ as a diagonal matrix. The patch radius of the uniform distribution was set as 0.05. Examples of the observed data generated from the normal and uniform distributions are shown in Fig. 4.

For the EMA agent, the discount rate is set to $\alpha = 0.00005$. The initial value of the estimate was set as $\mu^0 = \begin{pmatrix} 0.0 \\ 0.0 \end{pmatrix}$. For the parameters of the non-Min agent using the BDI, the number of hypotheses was set to one, and the discount rate was set to $\alpha = 0.0001$. The upper limit of the modification used by non-Min was set as $\Delta_{max} = 0.01$. For the initial values of the hypothesized model parameters, the initial estimated values were set to $\mu^0 = \begin{pmatrix} 0.0 \\ 0.0 \end{pmatrix}$, as in the EMA. The covariance matrix was fixed at $\Sigma = \begin{pmatrix} 0.05 & 0 \\ 0 & 0.05 \end{pmatrix}$, which is a diagonal matrix that is time-invariant. The likelihood contours shown in Fig. 3 are concentric circles when the covariance matrix is established. Therefore, bringing the estimate closer to the observed data at the shortest distance, as in EMA, is equivalent to modifying the estimate to minimize the amount of movement, as in the steepest descent method.

The data-generating distribution and hypothetical model are different even if they have the same normal distribution. The data-generating distribution was utilized to produce the observed data, whereas the hypothetical model was the distribution utilized as the basis for Bayesian inference.

## III. RESULTS

### 3.1 Simulation results

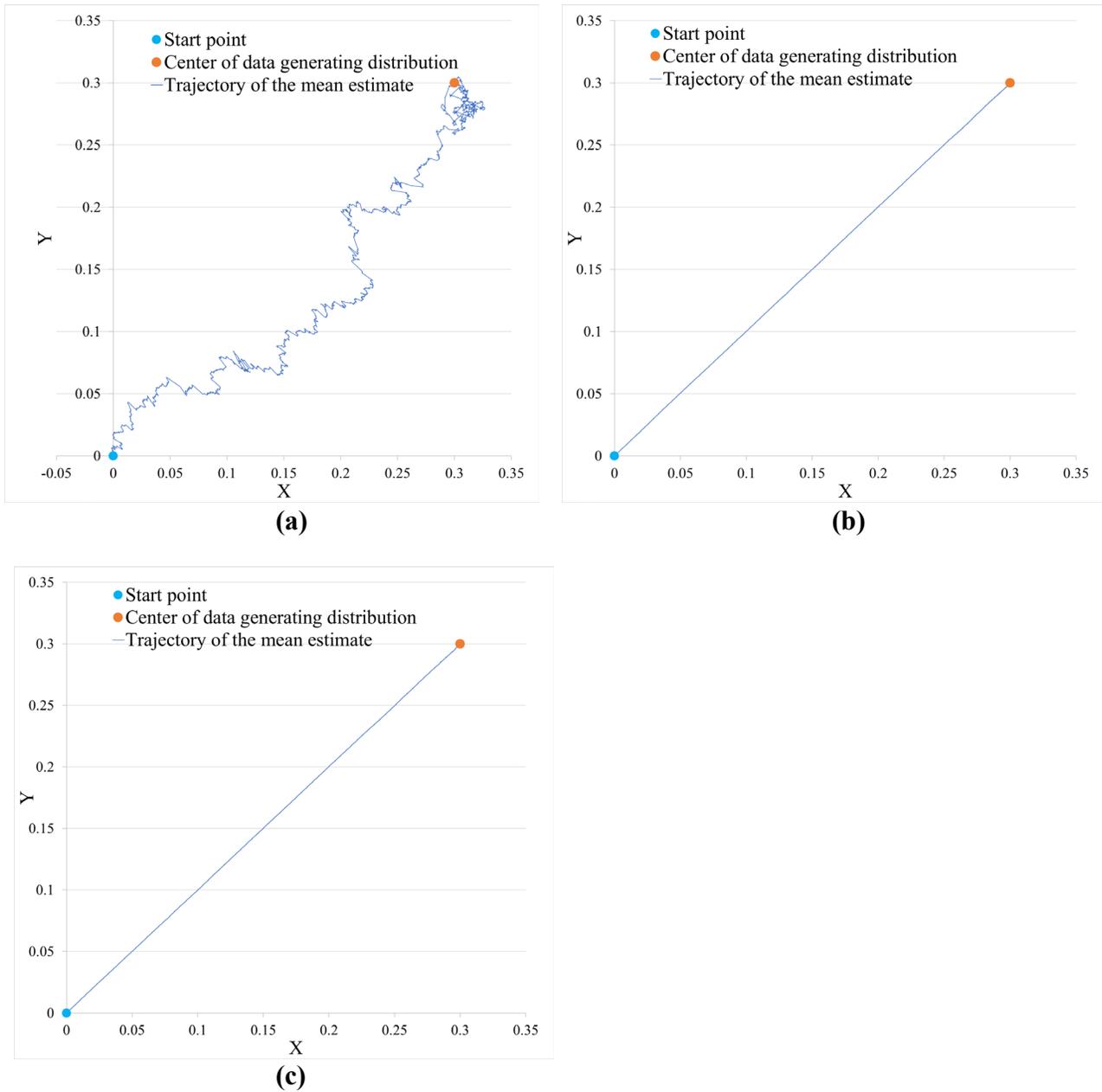

**Figure 5.** Trajectory of estimates for each agent from zero to 100,000 steps (for normal distribution). (a) non-Min agent, (b) Min agent, and (c) EMA agent.

Figures 5, 6, and 7 show the results for the EMA and two BDI agents, namely non-Min and Min agents, when sampling data from a normal distribution. Figure 5 shows the trajectory of the estimates for each agent from the time of 0 to 100,000 steps. In these figures, a straight line connects the two time-adjacent estimated

coordinates $\mu^{t-1}$ and $\mu^t$. Each agent begins at point $\begin{pmatrix} 0.0 \\ 0.0 \end{pmatrix}$ and gradually approaches point $\begin{pmatrix} 0.3 \\ 0.3 \end{pmatrix}$, which is the central coordinate of the data-generating distribution.

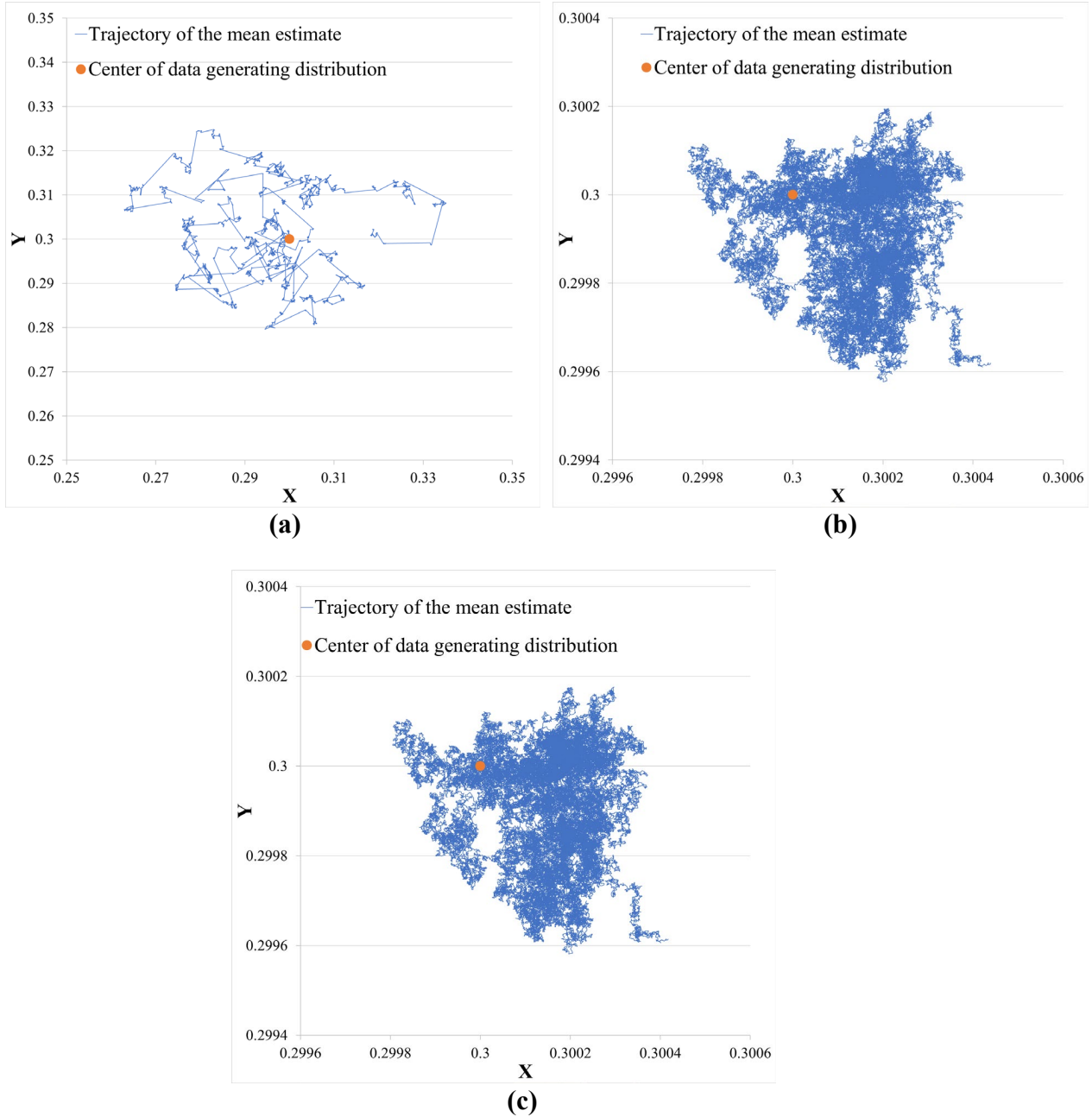

**Figure 6. Trajectory of estimates for each agent from 200,000 to 300,000 steps after convergence (for a normal distribution). (a) non-Min agent, (b) Min agent, and (c) EMA agent.**

Figure 6 shows the trajectory of the estimates from the time of 200,000 to 300,000 steps after each agent's estimates converged and reached a steady state near the center of the data-generating distribution.

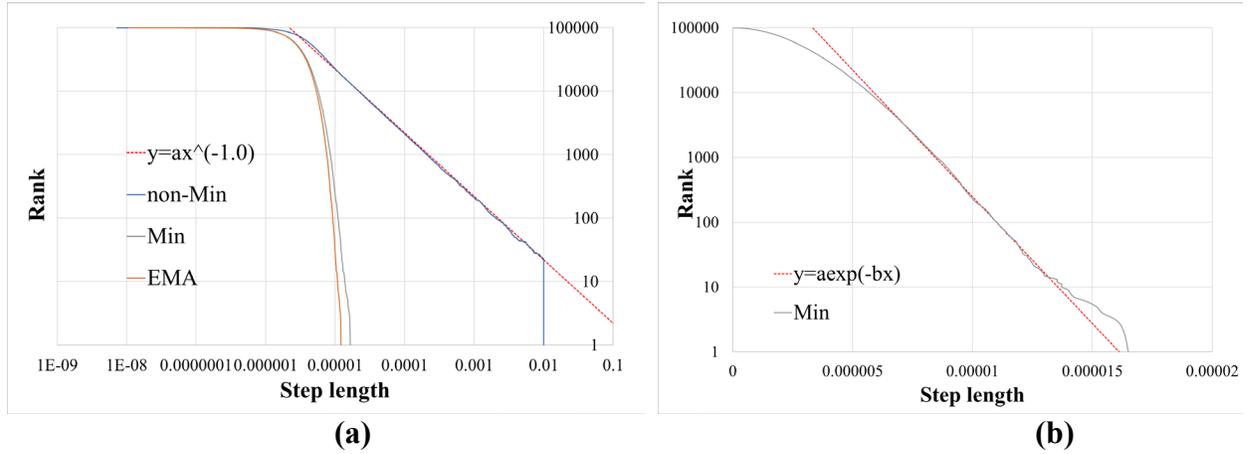

**Figure 7. Step length distribution over the 200,000–300,000 steps period (for a normal distribution). (a) Each agent. Both logarithmic displays and (b) Min agent. Only the vertical axis is shown in logarithms.**

Figure 7 shows the step length $l^t = \|\mu^t - \mu^{t-1}\|$, which represents the distance between adjacent estimated coordinates on the horizontal axis, and the ranking of step lengths in descending order on the vertical axis. Both axes are shown on a logarithmic scale in Fig. 7 (a). Because the rank of a step length represents the number of step lengths greater than or equal to the step length, this figure represents a complementary cumulative distribution function (CCDF). Figure 7(a) shows the data for each agent and the exponent one power-law distribution with a dashed line. This figure shows that the CCDF of step lengths for the non-Min agent is a power-law distribution with an exponent of approximately 1.0 for a wide range of step length data. This indicates that the frequency distribution of step lengths is $P(l) \sim l^{-2}$ or the Lévy walk with an exponent of two. The maximum amount of movement is limited to $\Delta_{max} = 0.01$, thereby no longer step lengths can appear.

The distributions of the Min and EMA agents are not power laws. Figure 7(b) shows the results for the Min agent on a logarithmic graph. The figure shows an exponential distribution with a dashed line. Additionally, the CCDF of the Min agent's step length can be approximated as an exponential distribution with respect to the high-rank step length. This indicates that the trajectory of the Min agent is not a Lévy walk, but a Brownian walk.

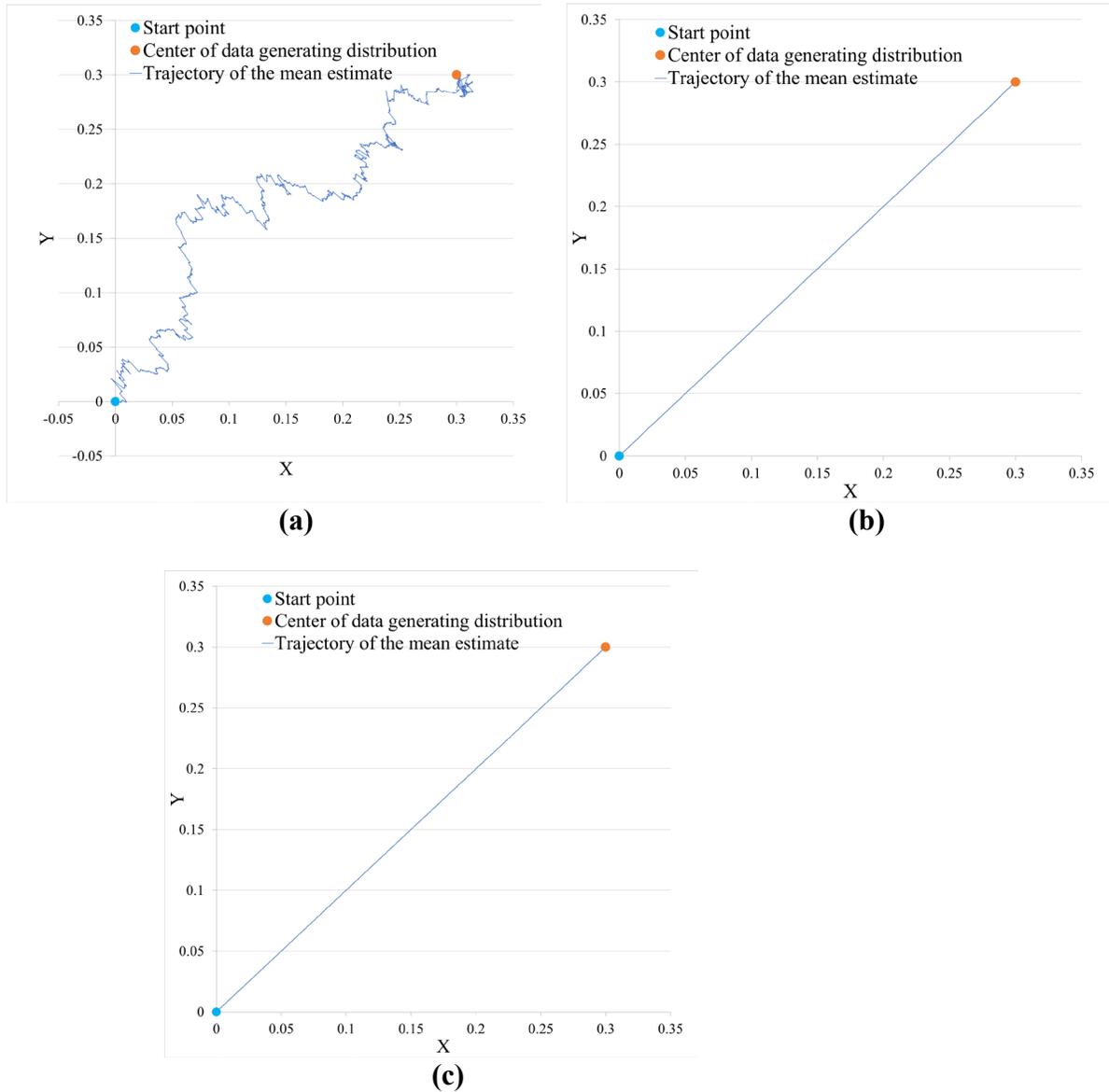

**Figure 8. Trajectory of estimates for each agent from zero to 100,000 steps (for a circular uniform distribution). (a) non-Min agent, (b) Min agent, and (c) EMA agent.**

Figures 8, 9, and 10 show the results when the observed data were sampled from a circular uniform

distribution with a radius of 0.05. Figure 8 shows the trajectory of each agent's estimate from time 0 to 100,000 steps, as shown in Fig. 5. As in the case of a normal distribution, the estimates for each agent approach the center coordinates of the generative distribution over time. Figure 9 shows the trajectory of the estimates from time 200,000 to 300,000 steps, as shown in Fig. 6. Figure 10(a) shows the distribution of step lengths for each agent, as in Fig. 7(a). This figure shows that the trajectory of the non-Min agent is a Lévy walk with an exponent of two, even when the data-generating distribution is uniform. The distributions of the Min and EMA agents were practically identical and did not exhibit a power-law distribution. Figure 10(b) shows the results for the Min agent. This figure is not displayed logarithmically on either the vertical or horizontal axis, and a straight line is shown as a dashed line. This figure shows that the CCDF of the Min agents' step lengths can be approximated linearly with respect to step lengths with a high rank.

Thus, the trajectory of the estimates for the non-Min agent is the Cauchy walk, regardless of the form of the data-generating distribution. However, for the EMA and Min agents, the characteristics of the step-length distributions changed based on the data-generating distribution. However, they did not become Lévy walks.

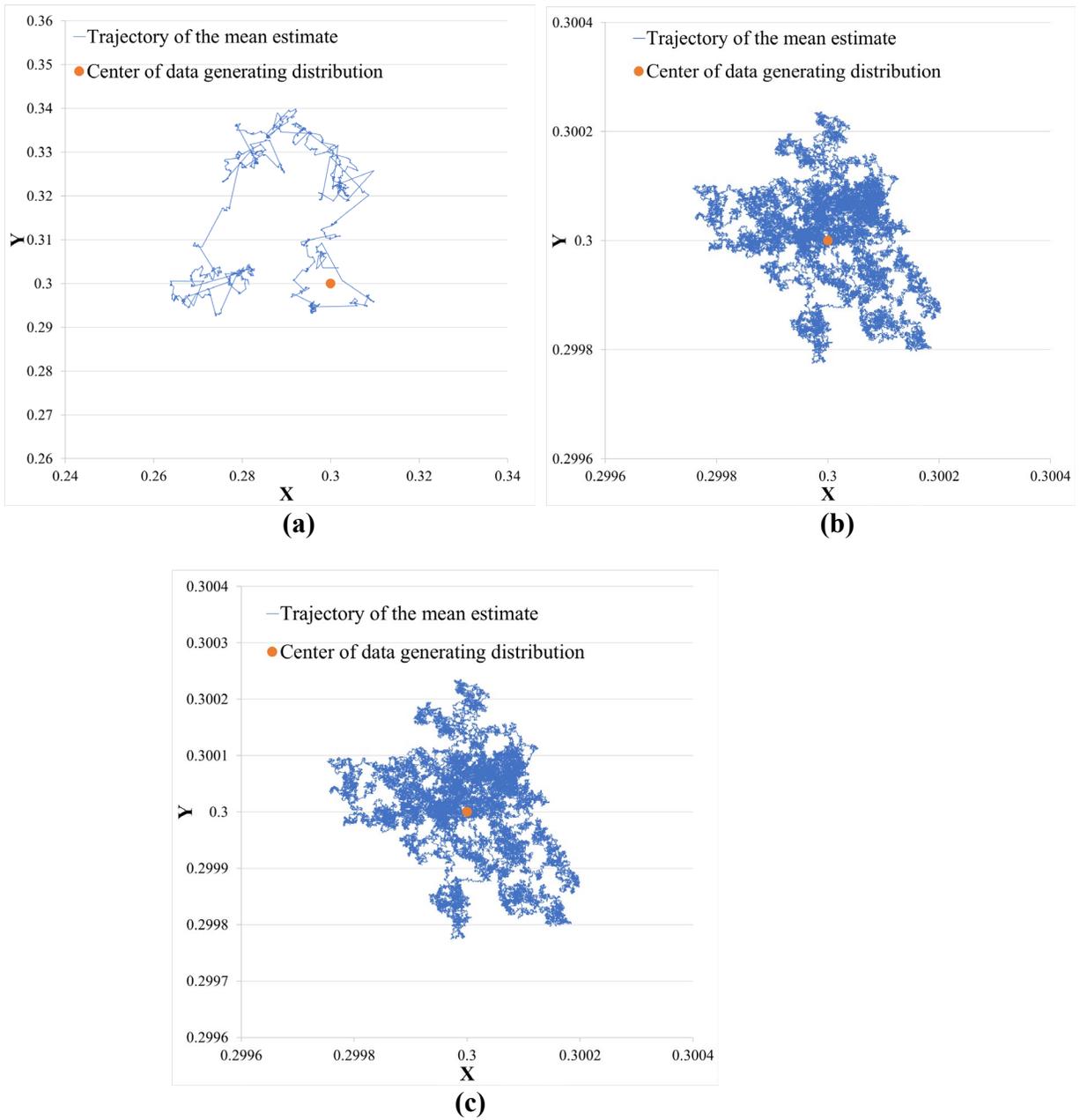

**Figure 9. Trajectory of estimates for each agent from 200,000 to 300,000 steps after convergence (for a circular uniform distribution). (a) non-Min agent, (b) Min agent, and (c) EMA agent.**

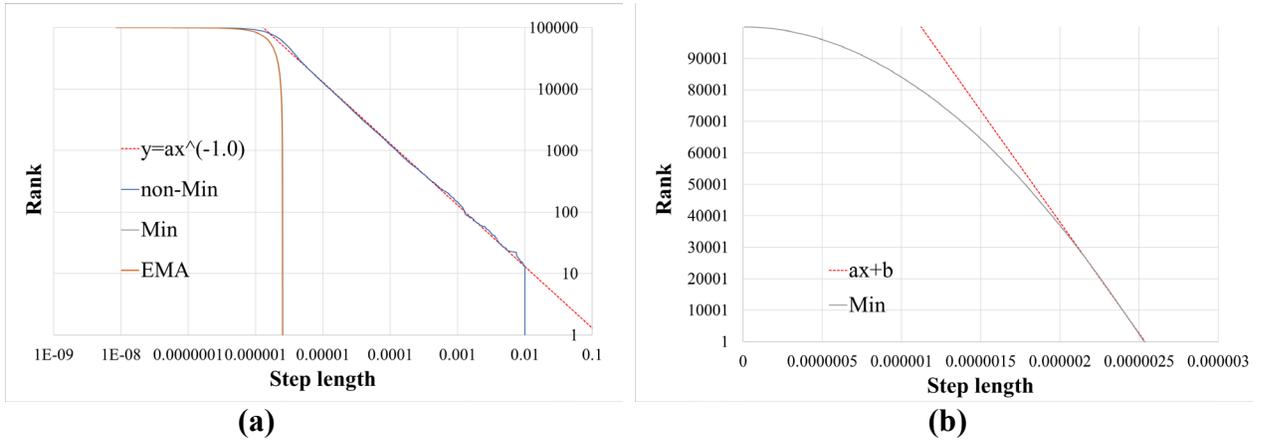

**Figure 10. Step-length distribution over the 200,000–300,000 period (for a circular uniform distribution) (a) Each agent. Both logarithmic displays and (b) Min agent.**

## 3.2 Lévy walk analysis

We analyze the reasons for the appearance of Lévy walks with exponent two in the non-Min agent. The step lengths in each direction are denoted as $|\Delta \mu_x| = l_x$ and $|\Delta \mu_y| = l_y$, respectively. The step length is denoted by $l = \sqrt{l_x^2 + l_y^2}$. The difference vector between the observed data and mean estimate is denoted by

$$\mu - d = \begin{pmatrix} \mu_x - d_x \\ \mu_y - d_y \end{pmatrix} = \begin{pmatrix} r_x \\ r_y \end{pmatrix}.$$ The norm of the vector is denoted by $r = \sqrt{(r_x)^2 + (r_y)^2}$.

In a non-Min agent, the allocation ratio of the modification amount in the *x*- and *y*-axis directions is $\beta$ and $1-\beta$, respectively. The step length $l_x$ in the *x*-direction can be written as follows using Eq. (23):

$$l_x \approx \left|\beta\Delta Z \frac{\partial \mu_x}{\partial Z}\right|$$

$$= \left|\beta\alpha\left[1-\exp\left(-\frac{(r_x)^2+(r_y)^2}{2\sigma^2}\right)\right]\frac{\sigma^2}{-r_x \exp\left(-\frac{(r_x)^2+(r_y)^2}{2\sigma^2}\right)}\right|. \quad (26)$$

$$= \left|\beta\alpha\sigma^2\left[1-\exp\left(\frac{r^2}{2\sigma^2}\right)\right]\frac{1}{r_x}\right|$$

$$= \beta\alpha\sigma^2\left(\exp\left(\frac{r^2}{2\sigma^2}\right)-1\right)\frac{1}{|r_x|}$$

Similarly, $l_y \approx (1-\beta)\alpha\sigma^2\left(\exp\left(\frac{r^2}{2\sigma^2}\right)-1\right)\frac{1}{|r_y|}$ holds for $l_y$.

If $r_x = r\cos\theta$ and $r_y = r\sin\theta$, the step length at a point with radius $r$ and angle $\theta$ with a horizontal axis can be written as

$$l = \sqrt{l_x^2 + l_y^2}$$

$$\approx \sqrt{\left(\beta\alpha\sigma^2\left(\exp\left(\frac{r^2}{2\sigma^2}\right)-1\right)\frac{1}{|r_x|}\right)^2 + \left((1-\beta)\alpha\sigma^2\left(\exp\left(\frac{r^2}{2\sigma^2}\right)-1\right)\frac{1}{|r_y|}\right)^2}$$

$$= \alpha\sigma^2\left(\exp\left(\frac{r^2}{2\sigma^2}\right)-1\right)\sqrt{\frac{\beta^2}{r^2(\cos\theta)^2}+\frac{(1-\beta)^2}{r^2(\sin\theta)^2}}, \quad (27)$$

$$= \alpha\sigma^2\frac{\left(\exp\left(\frac{r^2}{2\sigma^2}\right)-1\right)}{r}\sqrt{\frac{\beta^2}{(\cos\theta)^2}+\frac{(1-\beta)^2}{(\sin\theta)^2}}$$

$$= \alpha\sigma^2 C_r C_\theta$$

where $C_r = \dfrac{\left(\exp\left(\frac{r^2}{2\sigma^2}\right)-1\right)}{r}$ and $C_\theta = \sqrt{\dfrac{\beta^2}{(\cos\theta)^2}+\dfrac{(1-\beta)^2}{(\sin\theta)^2}}$, respectively.

Here from Eq. (27), $\left|\dfrac{\partial l}{\partial \theta}\right| \approx \alpha\sigma^2 C_r \left|\dfrac{\partial C_\theta}{\partial \theta}\right|$ and $\alpha\sigma^2 C_r \approx \dfrac{l}{C_\theta}$. Therefore,

$$\left|\frac{\partial l}{\partial \theta}\right| \approx \frac{l}{C_\theta}\left|\frac{\partial C_\theta}{\partial \theta}\right|. \tag{28}$$

Similarly, $\left|\frac{\partial l}{\partial r}\right| \approx \alpha\sigma^2 C_\theta \left|\frac{\partial C_r}{\partial r}\right|$ and $\alpha\sigma^2 C_\theta \approx \frac{l}{C_r}$. Therefore,

$$\left|\frac{\partial l}{\partial r}\right| \approx \frac{l}{C_r}\left|\frac{\partial C_r}{\partial r}\right|. \tag{29}$$

The probability of occurrence of step length $l$ at a point with radius $r$ and angle $\theta$ with a horizontal axis can be described as

$$\begin{aligned}
P(l \mid r, \theta) &\propto \left|\frac{\partial r}{\partial l}\right|\left|\frac{\partial \theta}{\partial l}\right| \\
&\approx \frac{C_r}{l}\left|\frac{\partial r}{\partial C_r}\right|\frac{C_\theta}{l}\left|\frac{\partial \theta}{\partial C_\theta}\right| \\
&= \frac{1}{l^2} C_r \left|\frac{\partial r}{\partial C_r}\right| C_\theta \left|\frac{\partial \theta}{\partial C_\theta}\right|
\end{aligned} \tag{30}$$

If the data-generating distribution is uniform and the probability of occurrence of the observed data at a point with radius $r$ and angle $\theta$ with the horizontal axis can be approximated by $P(r,\theta) \propto r\, d\theta\, dr$, then the probability of occurrence of $l$ can be approximated as

$$\begin{aligned}
P(l) &= \int_r \int_\theta P(l \mid r, \theta) P(r, \theta) \\
&\propto \int_r \int_\theta \frac{1}{l^2} C_r \left|\frac{\partial r}{\partial C_r}\right| C_\theta \left|\frac{\partial \theta}{\partial C_\theta}\right| r\, d\theta\, dr \\
&= \frac{1}{l^2} \int_r r C_r \left|\frac{\partial r}{\partial C_r}\right| \left(\int_\theta C_\theta \left|\frac{\partial \theta}{\partial C_\theta}\right| d\theta\right) dr \\
&\propto l^{-2}
\end{aligned} \tag{31}$$

Similarly, if the data-generating distribution is normal with the variance $\sigma_d^2$ and the probability of occurrence of the observed data at a point with radius $r$ and angle $\theta$ with the horizontal axis can be approximated by $P(r,\theta) \propto r\exp\left(-\dfrac{r^2}{2\sigma_d^2}\right)d\theta dr$, then the probability of occurrence of $l$ can be approximated as

$$\begin{aligned}
P(l) &= \int_r \int_\theta P(l \mid r,\theta) P(r,\theta) \\
&\propto \int_r \int_\theta \dfrac{C_r}{l}\left|\dfrac{\partial r}{\partial C_r}\right|\dfrac{C_\theta}{l}\left|\dfrac{\partial \theta}{\partial C_\theta}\right| r \exp\left(-\dfrac{r^2}{2\sigma_d^2}\right) d\theta dr \\
&= \dfrac{1}{l^2} \int_r r\exp\left(-\dfrac{r^2}{2\sigma_d^2}\right) C_r \left|\dfrac{\partial r}{\partial C_r}\right| \left(\int_\theta C_\theta \left|\dfrac{\partial \theta}{\partial C_\theta}\right| d\theta\right) dr \\
&\propto l^{-2}
\end{aligned} \qquad (32)$$

Thus, the occurrence probability of the step length can be approximated as a power-law distribution with exponent two.

## IV. DISCUSSION

The main objective of this study was to propose a simple model that universally generates Cauchy walks and to identify the conditions under which Cauchy walks appear. We performed an online estimation simulation to estimate the center of the data-generating distribution based on the observed data. The results revealed that among the two agents using BID, the non-Min agent showed a Cauchy walk in the estimation behavior, regardless of whether the data-generating distribution was normal or uniform. In the case of the Min and EMA agents, the frequency distribution of step length occurrence depends on the form of the data-generating distribution and is not a Lévy walk. The estimated behavior of the Min agent was obtained as a Brownian walk when the data-generating distribution was normal.

The two agents using a BID share the same objective of modifying the mean estimates to increase the probability of occurrence of the observed data, as shown in Eqs. (16) and (17). However, the means of achieving this goal differ.

Similar to the EMA agent, the Min agent uses a strategy of moving estimates closer to the observed data at the shortest distance. The non-Min agent weakens the shortest-distance constraint and moves the estimate closer to the observed data. As a matter of course, step lengths are longer for the non-Min agent than for the Min agent. The analysis revealed that Cauchy walks with exponent two appear for the non-Min agent.

Consider a strategy for moving the estimates closer to the observed data. If the estimates are far from the center of the data-generating distribution, moving the estimates closer to the observed data generated near the center will move them closer to the center of the generating distribution, which is the true value. Moreover, in this situation, the location of the observed data is a crucial indication of the central coordinates to be searched, and bringing the estimated value close to the observed data is congruent with the original purpose of estimating the central coordinates from the observed data. In this case, the two agents using BID and EMA approach the central coordinates.

In contrast, in a situation where the estimated values are sufficiently close to the center coordinates, or in an extreme case where the estimated values coincide with the center, moving the estimated values closer to the observed data would move the estimated values away from the center. Moreover, in steady-state situations where the estimates converge near the center, the observed data are not useful for estimating the center coordinates. Thus, the information in the observed data varies based on the positional relationship between the estimated and center coordinates, that is, the scale of the distance between them. In the steady state after convergence, the observed data are no longer meaningful information, and moving the estimates closer to the observed data creates a random walk. However, the same random walk differs in that it becomes a Lévy walk for the non-Min agent and a Brownian walk for the Min agent.

In the LFFH, the conditions that make Lévy walk an optimal foraging behavior are the sparsity of food and a lack of information (memory) about the predator's food [14]. Without any indication of prey, the predator must conduct a random search. To answer the question of why Lévy walks appear in such a situation instead of the Brownian walk, we should understand why the Lévy walk is spontaneously produced, independent of external input stimuli.

Abe found that Lévy walks arise from critical phenomena in systems such as the brain and have functional advantages in information processing [20]. However, Abe's model did not generate a Lévy walk with an exponent of two. In the LFFH, the advantage of the most efficient search has been cited as the reason for the appearance of Lévy walks with exponent two. However, this is not necessarily true for spaces larger than two dimensions [15]. Additionally, the simulations in this study reveal that the behavior of the non-Min agent exhibits a Lévy walk with exponent two in the steady state when the observed data are no longer useful as information because the non-Min agent does not have the means or ability to detect the shortest path, unlike the EMA and Min agents.

We consider the implications of minimizing the modification of the estimates used by the EMA and Min agents. In contrast, bringing the estimates closer to the observed data using Eq. (1) implies that the information in the observed data is reflected in the next estimate. However, as shown in Eqs. (2) and (3), the current estimates incorporate information from past observational data. The minimum strategy involves incorporating information from new observation data, while minimizing the loss of information from past observation data. Therefore, this strategy is the most efficient way to incorporate information from observed data into estimates.

Conversely, the non-Min agent, who brings their estimates closer to the observed data and does not use a minimum strategy, incorporates new information from the observed data while simultaneously discarding past information. However, the difference between this and the larger discount rate should be considered. As the EMA and Min agents continue to incorporate new information from the observed data, fluctuations occur owing to the randomness of the observed data. The larger the discount rate, the larger the fluctuation. Its noise characteristics depend on the form of the data-generating distribution, which, in the case of a normal distribution, results in a Brownian walk. Even if the discount rate for the non-Min agent is the same as for EMAs and Mins, apart from the external fluctuations resulting from the observed data, an internal fluctuation exists because it does not use the minimum strategy, which is the Lévy walk, regardless of the distribution generating the data.

In the future, the significance of the minimum strategy and its weakening on the foraging behavior and estimation process of organisms should be clarified. In the simulations in this study, a normal distribution was employed as the basis used by the BID agent for the estimation. However, its validity must be verified. The covariance matrix of the basis was fixed as $\Sigma = \begin{pmatrix} 0.05 & 0 \\ 0 & 0.05 \end{pmatrix}$ for the simulation. Smaller variances render distant data undetectable, whereas larger variances result in less-accurate estimates. The effect of this difference in variance on the behavior of the agents should be examined. These issues should be addressed in future research.

# Funding

This work was supported by JSPS KAKENHI [grant number JP21K12009].

# Data Availability

The source code and data used to produce the results and analyses presented in this manuscript are available from the GitHub repository:

https://github.com/Kaga-1945/SinoharaLabo_code.git

38# Figure Legends

Figure 1. Online estimation task overview. The agent observes randomly generated data based on a certain probability distribution one by one each time and successively estimates the central coordinates of the distribution.

Figure 2. Overview of EMA. (a) EMA is an algorithm that moves the estimate close to the observed data. (b) If the goal is simply to reduce the distance between the estimate and the data by , the estimate can be moved anywhere on the one inner concentric circle. Conversely, EMA is an algorithm that moves the estimate close to the data at the shortest distance.

Figure 3. Likelihood contours. If we want to increase the likelihood by , we can move the estimate   to any position on the inner contour. Moreover, an infinite number of possible solutions are presented.

Figure 4. Examples of observed data. (a) When the data-generating distribution is a normal distribution. (b) When the data-generating distribution is a circular uniform distribution.

Figure 5.   Trajectory of estimates for each agent from zero to 100,000 steps (for normal distribution). (a) non-Min agent, (b) Min agent, and (c) EMA agent.

Figure 6. Trajectory of estimates for each agent from 200,000 to 300,000 steps after convergence (for a normal distribution). (a) non-Min agent, (b) Min agent, and (c) EMA agent.

Figure 7. Step length distribution over the 200,000–300,000 steps period (for a normal distribution). (a) Each agent. Both logarithmic displays and (b) Min agent. Only the vertical axis is shown in logarithms.

39Figure 8. Trajectory of estimates for each agent from zero to 100,000 steps (for a circular uniform distribution). (a) non-Min agent, (b) Min agent, and (c) EMA agent.

Figure 9. Trajectory of estimates for each agent from 200,000 to 300,000 steps after convergence (for a circular uniform distribution). (a) non-Min agent, (b) Min agent, and (c) EMA agent.

Figure 10. Step-length distribution over the 200,000–300,000 period (for a circular uniform distribution) (a) Each agent. Both logarithmic displays and (b) Min agent.